\documentclass[twocolumn,showpacs,preprintnumbers,prl,amsmath,amssymb]{revtex4}
\usepackage{epsfig}
\usepackage{bm} 
\begin{document}
\draft
\title{Non-Fermi liquid behavior of quasi one-dimensional pseudogap materials}
\author{Lorenz Bartosch}

\affiliation{Institut f\"{u}r Theoretische Physik, Universit\"{a}t Frankfurt,
Robert-Mayer-Strasse 8, 60054 Frankfurt, Germany}
\date{August 26, 2002}
\begin{abstract}
We study the spectral function of the pseudogap phase of quasi one-dimensional charge density wave materials. 
Using a stochastic approach and emphasizing an exact treatment of non-Gaussian order parameter fluctuations we will go beyond a usual perturbative calculation.
Our results give a good fit to ARPES data and explain the absence of the Fermi edge in charge density wave materials even above the Peierls transition, indicating non-Fermi liquid behavior.
\end{abstract}
%
%
%
%
\maketitle



Non-Fermi liquid behavior is a central topic of modern condensed matter physics.
A paradigmatic example represents the universality class of Luttinger liquids \cite{Haldane81,Voit94} whose spectral properties have been calculated exactly \cite{Meden92} and which applies to the low-energy physics of a variety of one-dimensional ($1d$) gapless systems.
However, recent {\em angle-resolved photoemission spectroscopy} (ARPES) measurements of charge density wave materials like the {\em blue bronze} $\text{K}_{0.3} \text{Mo} \text{O}_3$ or $(\text{Ta}\, \text{Se}_4)_2\,\text{I}$ \cite{Dardel91,Claessen95,Gweon96,Gweon01,Perfetti02} show that these materials cannot be understood in terms of Luttinger liquid theory, even at temperatures far above the Peierls transition. 
It turns out that the 
pseudogap observed in the charge density wave materials is similar to that found in the underdoped phase of high temperature superconductors \cite{Loeser96,Damascelli02}.
It is this analogy which makes a better understanding of the $1d$ pseudogap phase even more interesting.

The relative proximity of the pseudogap phase of charge density wave materials to the Peierls transition indicates that strong electron-phonon interactions should play a dominant role.
In the early seventies, Lee, Rice, and Anderson \cite{Lee73} studied the single-particle Green function of a $1d$ model based on order parameter fluctuations 
within the second-order Born approximation.
Although this calculation could qualitatively explain the observed pseudogap phenomenon \cite{Shannon00}, its validity hinges on its perturbative nature.
As is well known for correlated $1d$ electronic systems, interference effects due to perfect nesting between the two Fermi points invalidate a perturbative calculation near the Fermi level. Therefore, one should rather try to find a more accurate or even exact solution. A widely used partial fraction expansion of the single-particle Green function presented by Sadovskii \cite{Sadovskii79} was intended to be exact for Gaussian order parameter fluctuations but turned out to be only an approximate solution to the problem raised by Lee, Rice, and Anderson \cite{Tchernyshyov99}.

Numerical simulations based on exact diagonalization methods \cite{Millis00b} or the phase formalism \cite{Bartosch99d,Bartosch00d} gave exact results for the density of states (DOS) for Gaussian order parameter fluctuations but showed also that Gaussian order parameter fluctuations are inapplicable at low temperatures where amplitude fluctuations should get frozen out and a hard gap should appear in the limit $T \to 0$ if quantum fluctuations are neglected. 
As pointed out in Ref.\ \onlinecite{Bartosch00d}, the right low-temperature physics can be described within a model based on classical phase fluctuations.
Using the phase formalism outlined in Ref.\ \onlinecite{Bartosch00d}, the DOS was also calculated numerically for the crossover regime to non-Gaussian order parameter fluctuations using a standard Ginzburg-Landau theory \cite{Monien01}.
How to find non-perturbative results for the complete spectral function which could be compared with ARPES data,
however, remained an open question.

In this work we shall turn to this problem and calculate the spectral function of the $1d$ pseudogap phase of charge density wave materials for order parameter fluctuations given by a generalized Ginzburg-Landau functional.
We generalize a stochastic method originated by Halperin \cite{Halperin65,Lifshits88} for a particle in a white noise potential towards pseudogap systems involving disorder of the colored noise type.

As a starting point, we use the fluctuating gap model \cite{Lee73,Sadovskii79,Lifshits88} 
which couples 
the electronic degrees of freedom 
to a complex order parameter field $\tilde\Delta(x) \equiv \Delta(x) e^{i\vartheta(x)}$ with amplitude $\Delta(x)$ and phase $\vartheta(x)$. The two-component wave functions $\bm{\psi}_n(x)$ satisfy a Dirac-like equation \footnote{In this work we use units such that $\hbar = v_F = k_B = 1$. Later we will reintroduce the Fermi velocity $v_F$.},
\begin{equation}
  \label{eq:Dirac}
  [-i\bm{\sigma}_3 \partial_x + \tilde\Delta(x)  \bm{\sigma}_{+} + \tilde\Delta^{\ast}(x) \bm{\sigma}_{-} ] \bm{\psi}_n(x) = \omega_n \bm{\psi}_n(x) \;. 
\end{equation}
Here $\bm{\sigma}_{\pm} \equiv (\bm{\sigma}_{1} \pm i \bm{\sigma}_{2})/2$ and $\bm{\sigma}_3$ are the usual Pauli matrices.
Since the Hamiltonian corresponding to Eq.\ (\ref{eq:Dirac}) is form invariant under the charge conjugation operation $\bm{H} \to \bm{\sigma}_{1} \bm{H}^{\ast} \bm{\sigma}_{1}$ it is possible to use boundary conditions which imply that all eigenstates $\bm{\psi}_n(x)$ are invariant under the charge conjugation operation $\bm{\psi}_n(x) \to \bm{\sigma}_{1} \bm{\psi}_n^{\ast}(x)$. 
All these eigenstates may be found by making the Eikonal ansatz
\begin{equation}
  \label{eq:wavefunction}
  \bm{\psi}(x) = \left(
    \begin{array}{c}
      \psi(x) \\
      \psi^{\ast}(x) \\
    \end{array} \right) =
A e^{\zeta(x)}\left(
    \begin{array}{c}
      i e^{i\varphi(x)/2} \\
      -i e^{-i\varphi(x)/2} \\
    \end{array} \right)\;.
\end{equation}
In the thermodynamic limit $L \to \infty$, where $L$ is the length of the chain,
all physical quantities we will be interested in are independent of these boundary conditions \cite{Lifshits88}.
The logarithmic amplitude $\zeta(x)$ and the phase $\varphi(x)$ satisfy the following equations of motion \cite{Bartosch00d}, 
\begin{subequations}
  \label{eq:motion}
\begin{align}
  \label{eq:phi}
  \partial_x \varphi(x) & = 2 \omega + 2 \Delta(x) \cos\left[\varphi(x)-\vartheta(x)\right] \;, \\
  \label{eq:zeta}
  \partial_x \zeta(x) & = \Delta(x) \sin\left[\varphi(x)-\vartheta(x)\right] \;.
\end{align}
\end{subequations}
$\varphi(x)$ and $\zeta(x)$ are simply related to the DOS $\rho(\omega)$ and the inverse localization length (or Lyapunov exponent) $\ell^{-1}(\omega)$ by \cite{Bartosch00d}
\begin{equation}
  \rho(\omega) = \displaystyle{\frac{1}{2\pi}} \partial_\omega \langle \partial_x \varphi(x) \rangle \;,\;
  \ell^{-1}(\omega) = \langle \partial_x \zeta(x) \rangle \;.
\end{equation}
Here and in the following, $\langle \dots \rangle$ indicates averaging with respect to the probability distribution of $\tilde \Delta(x)$ which we will specify below.

The spectral function for right- or left-moving electrons ($\alpha = \pm 1$)
is defined in terms of the appropriately normalized wave functions $\bm{\psi}_{n}(x)$ as
\begin{align}
  & A^{\alpha}(k;\omega)\equiv A (\alpha (k_F+k);\omega) \nonumber \\
  & \quad \equiv \frac{1}{L} \left\langle \sum_{n}  \left|\int_{-L/2}^{L/2} dx\, e^{-ikx} \psi_{n}(x)\right|^2 \delta(\omega-\omega_n) \right\rangle\;.
  \label{eq:spectralfunction}
\end{align}
Note that we measure the deviation $k$ of a wave vector $p = \alpha(k_F + k)$ from one of the two Fermi wave vectors $\pm k_F$ locally outwards such that due to inversion symmetry $A^{\alpha}(k;\omega)$ is independent of $\alpha$.

Instead of using $\int |\bm{\psi}_n|^2\, dx = 1$ to normalize the wave functions let us now use the normalization condition $|\psi_n(0)|=1$.
Since the process of averaging restores translational invariance, it follows
\begin{equation}
  A^{\alpha}(k;\omega) 
  = \left\langle \sum_{n} \frac{\int_{-L/2}^{L/2} dx\, e^{-ikx} \psi_{n}(x)\psi_{n}^{\ast}(0)}{2 \int_{-L/2}^{L/2} dx\, \left|\psi_{n}(x)\right|^2}\, \delta(\omega-\omega_n) \right\rangle\;.
  \label{eq:spectralfunctionII}
\end{equation}

Of course, not every solution to Eqs.\ (\ref{eq:wavefunction}) and (\ref{eq:motion}) is an eigenstate of the system. To construct the eigenstates, we follow Ref.\ \onlinecite{Lifshits88} and integrate the equations of motion (\ref{eq:motion}) from the boundaries at $-L/2$ and $L/2$ to $x=0$ where we try to match these solutions. Let us call the solutions to Eqs.\ (\ref{eq:phi}) and (\ref{eq:zeta}) when advancing the solution from $\mp L/2$ to the right (left) $\varphi_{\pm}(x)$ and $\zeta_{\pm}(x)$.
$\psi(x)$ is then given by
\begin{equation}
  \label{eq:psiplus}
  \psi (x) = \left\{
    \begin{array}{ll}
     i e^{-\zeta_{+}(0) + \zeta_{+}(x) + i\varphi_{+}(x)/2} \;, \;x < 0 \\
     i e^{-\zeta_{-}(0) + \zeta_{-}(x) + i\varphi_{-}(x)/2} \;,\;x > 0
    \end{array} \right. \;.
\end{equation}
The eigenstates $\bm{\psi}_n(x)$ are determined by the energies $\omega_n$ for which the matching condition $\varphi_{+}(0)=\varphi_{-}(0)$ (up to multiples of $2\pi$) is satisfied. In this case $\varphi_{+}(x)=\varphi_{-}(x)$ for all $x$, and the eigenstates have their maxima somewhere in the bulk from where the wave functions fall off exponentially.

Let us also introduce
\begin{equation}
  \label{eq:U}
  U_{\pm}(x) \equiv \frac{e^{ikx}}{\psi(x)}\int_{\mp L/2}^{x} dx^{\prime} \, e^{-ikx^{\prime}} \psi(x^{\prime}) \;,
\end{equation}
which satisfy the additional equation of motion
\begin{equation}
  \label{eq:U}
  \partial_x U (x) = i[k-\omega-\Delta(x) e^{-i[\varphi(x)-\vartheta(x)]} ] U(x) + 1 \;.
\end{equation}
It is now possible to rewrite the spectral function as
\begin{align}
  A^{\alpha}(k;\omega) 
  & = \left\langle \left[U_{+}(0) - U_{-}(0) \right] \, \delta(\varphi_{+}(0)-\varphi_{-}(0)) \right\rangle \nonumber \\
  & = 2 \, \text{Re} \left\langle U_{+}(0) \, \delta(\varphi_{+}(0)-\varphi_{-}(0)) \right\rangle\;.
  \label{eq:spectralfunctionIII}
\end{align}
One may explicitly verify that the spectral function $A(k;\omega) = A^{\alpha}(|k|-k_F;\omega)$ satisfies the usual sum rules,
\begin{align}
  \int_{-\infty}^{\infty} d\omega \, A(k;\omega) & = 1 \;, \\
  \int_{-\infty}^{\infty} \frac{dk}{2\pi} \, A(k;\omega) & = 2 \langle \delta(\varphi_{+}(0) - \varphi_{-}(0) ) \rangle =  \rho(\omega) \;.
\end{align}



The order parameter fluctuations of $\tilde \Delta (x) = \Delta(x) e^{i\vartheta(x)}$ are determined by a free energy functional which has been derived in the context of superconductivity \cite{Kosztin98}. Expanding the free energy functional in powers of the order parameter field $\tilde \Delta(x)$ and derivatives thereof one obtains the usual Ginzburg-Landau functional. A better approximation which is not restricted to small $\Delta$ and hence is valid at arbitrary temperatures can be obtained by expanding only in gradients of the order parameter field. The {\em generalized} Ginzburg-Landau free energy functional including all terms up to second order in the gradients of the phase and the amplitude of the order parameter has the form \cite{Kosztin98}
\begin{align}
  \label{eq:Ginzburg_Landau}
  \beta F \{\tilde \Delta\} & =  \int_{-L/2}^{L/2} dx \, \left[ \frac{M(\Delta(x))}{2} \left(\partial_x \Delta(x)\right)^2 \right. \nonumber \\
  & \left. {} + \frac{n(\Delta(x))}{2} \left(\partial_x \vartheta (x)\right)^2 + V(\Delta(x)) \right] \;,
\end{align}
where
\begin{subequations}
\label{eq:MnV}
\begin{align}
  \label{eq:M}
  M(\Delta) & = \sum_{\tilde \omega_n > 0} \frac{\tilde \omega_n^2}{(\tilde \omega_n^2 + \Delta^2)^{5/2}} \;,\\
  n(\Delta) & = \sum_{\tilde \omega_n > 0} \frac{\Delta^2}{(\tilde \omega_n^2 + \Delta^2)^{3/2}} \;,\\
  V(\Delta) & = - 4 \!\!\!\! \sum_{0<\tilde \omega_n \lesssim \epsilon_0} \!\! \left( \sqrt{\tilde \omega_n^2 + \Delta^2} - \tilde \omega_n \right) + \frac{\Delta^2}{\pi \lambda T} \;.
  \label{eq:V}
\end{align}
\end{subequations}
Here, $T=1/\beta$ is the temperature of the system, $\lambda \ll 1$ is the dimensionless electron-phonon coupling constant, $\epsilon_0$ is an ultraviolet (bandwidth) cutoff, and $\tilde \omega_n = (2n+1)\pi T$ are fermionic Matsubara frequencies. In the weak coupling limit $\lambda \ll 1$, $V(\Delta)$ and hence $F \{\tilde \Delta\}$ depend on $\lambda$ and $\epsilon_0$ only via the critical mean-field temperature $T_c^{\text{MF}} = 1.134\, \epsilon_0 \exp(-1/\lambda)$.

\begin{figure}[tb]
\begin{center}
\epsfig{file=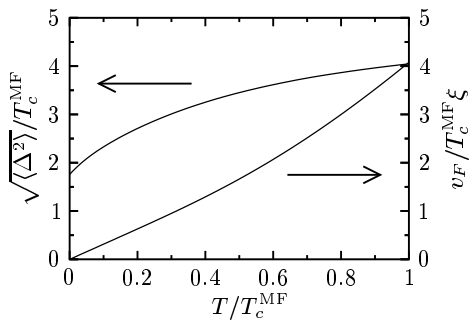} 
\end{center}
\vspace{-1mm}
\caption{Temperature dependence of $\sqrt{\langle \Delta^2(x) \rangle}$ and $\xi^{-1}$ for a lattice spacing $d = 0.2\, v_F/T_c^{\text{MF}}$ such that $\xi \gtrsim d$ for $T \lesssim T_c^{\text{MF}}$.}
\label{fig:correlationlength}
\end{figure}

The probability for a given realization of the order parameter field $\tilde \Delta(x)$ is proportional to $\exp(-\beta F \{\tilde \Delta\})$. This probability can be described as a stochastic process which due to the locality of $F \{\tilde \Delta\}$ has the Markov property. The probability $P(\Delta)$ that $\Delta(x)$ assumes the value $\Delta$ and the transition probabilities $T(\tilde \Delta(x) \! \gets \! \tilde \Delta(x^{\prime}))$ can be calculated using the transfer matrix method \cite{Scalapino72}. 
At this point we would like to note that the integral and derivatives in Eq.\ (\ref{eq:Ginzburg_Landau}) should be considered as a sum and finite differences. In the case of a generalized Ginzburg-Landau theory as considered here, $P(\Delta)$ and $T(\tilde \Delta(x) \gets \tilde \Delta(x^{\prime}))$ actually do depend on the lattice constant $d$.
In Fig.\ \ref{fig:correlationlength} we show $\sqrt{\langle \Delta^2 \rangle}$ and $\xi^{-1}$ as functions of temperature. For small temperatures we have $\xi \sim v_F/\pi T$ and $\sqrt{\langle \Delta^2 \rangle} \approx 1.76 \, T_c^{\text{MF}}$, consistent with the predictions of a strictly $1d$ theory based on classical phase fluctuations \cite{Gruener94,Bartosch00d}.
Since $\sqrt{\langle \Delta^2 \rangle} \gtrsim 1.76 \, T_c^{\text{MF}}$, our {\em generalized} Ginzburg-Landau functional given in Eqs. (\ref{eq:Ginzburg_Landau}) and (\ref{eq:MnV}) may not simply be replaced by a usual Ginzburg-Landau functional by expanding in $\Delta$.
In contrast to predictions based on standard Ginzburg-Landau theory as used in Refs.\ \onlinecite{Monien01} and \onlinecite{Scalapino72}, our calculations show that $\sqrt{\langle \Delta^2 \rangle}$ grows with increasing temperature.

The stochastic process corresponding to $T(\tilde \Delta(x) \gets\tilde \Delta(x^{\prime}))$ can be described by the Langevin equations
\begin{subequations}
\begin{align}
  \label{eq:Delta}
  \partial_x \Delta (x) & = a(\Delta)  + b(\Delta) \, \eta_{\Delta}(x) \;, \\
  \label{eq:theta}
  \partial_x \vartheta (x) & = \left(\frac{1}{n(\Delta(x))}\right)^{1/2} \, \eta_{\vartheta}(x) \;.
\end{align}
\end{subequations}
Here $\eta_{i}(x)$ with $i=\Delta,\vartheta$ represent Gaussian white noise with $\langle \eta_{i}(x)\eta_{j}(x^{\prime})\rangle=\delta_{i,j} \delta(x-x^{\prime})$.
$a(\Delta)$, $b(\Delta)$ and $P(\Delta)$ are simply related to 
$T(\Delta \! \gets \! \Delta^{\prime})$
and satisfy $a(\Delta) = b(\Delta){\partial_{\Delta} (b(\Delta) P(\Delta))}/{2 P(\Delta)}$.

To make further progress we absorb the phase of the order parameter field $\vartheta(x)$ by the gauge transformation \cite{Brazovskii76,Bartosch00d}
$\varphi(x) \to \varphi(x) + \vartheta(x)$.
We now obtain the equations of motion
\begin{subequations}
\begin{align}
  \label{eq:phiII}
  \partial_x \varphi(x) & = 2 \omega -  \partial_x \vartheta(x) + 2 \Delta(x) \cos\left[\varphi(x)\right] \;, \\
  \label{eq:zetaII}
  \partial_x \zeta(x) & = \Delta(x) \sin\left[\varphi(x)\right] \;, \\
  \label{eq:UII}
  \partial_x U (x) & = i [k-\omega-\Delta(x) e^{-i\varphi(x)} ] U(x) + 1 \;.
\end{align}
\end{subequations}
Since the stochastic process for a given realization of the order parameter field has the Markov property we may write Eq.\ (\ref{eq:spectralfunctionIII}) as
\begin{align}
  A^{\alpha}(k;\omega) 
   = \, & 2 \, \text{Re} \int_0^{\infty} d\Delta \int_{0}^{2\pi} d\varphi \, \frac{1}{P(\Delta)} \nonumber \\
   \times  &  P_1(\Delta,\varphi;0)   P_0(\Delta,-\varphi;0)\;.
  \label{eq:spectralfunctionIV}
\end{align}
Here
\begin{subequations}
\begin{align}
  P_0(\Delta,\varphi;x) = & \,
  \langle \delta(\varphi-\varphi_{+}(x)) \delta(\Delta-\Delta(x)) \rangle \;, \\
  P_1(\Delta,\varphi;x) = & \,
  \langle U_{+}(x) \, \delta(\varphi - \varphi_{+}(x) \delta(\Delta-\Delta(x)) \rangle \;.
\end{align}
\end{subequations}
An expression similar to Eq.\ (\ref{eq:spectralfunctionIV}) (with $P_1$ replaced by $P_0$) does hold for the DOS.
Making use of the Gaussian white noise property of the variables $\eta_i(x)$, one may show that $P_0$ and $P_1$ obey (with $\partial_{\Delta}$ acting on everything to its right)
\begin{subequations}
\begin{align}
  \partial_x P_0 = \big[ & \frac{\partial_{\varphi}^2}{2n} - 2 \partial_{\varphi} (\omega + \Delta\cos \varphi) + \frac{(\partial_{\Delta} b)^2}{2}  - \partial_{\Delta} a \big] P_0 \;,
 \label{eq:FP:P0}\\
  \partial_x P_1 = \big[ & \frac{\partial_{\varphi}^2}{2n} - 2 (\omega + \Delta\cos \varphi) \partial_{\varphi} + i (k-\omega -\Delta e^{i\varphi}) \big. \nonumber \\
 \big. {} + & \frac{(\partial_{\Delta} b)^2}{2}  - \partial_{\Delta} a \big] P_1 + P_0 \;.  \label{eq:PDE:P1}
\end{align}
\end{subequations}
The stationary solutions to the $2d$ Fokker-Planck equation (\ref{eq:FP:P0}) and the $2d$ partial differential equation (\ref{eq:PDE:P1}) may be found numerically. We then obtain the spectral function $A^{\alpha}(k;\omega)$ (as well as the DOS and the inverse localization length) by simple quadrature.
The spectral function, the DOS and the inverse localization length are therefore simultaneously directly calculable without any approximations. There is no need to simulate a long chain as was done for the DOS (and the inverse localization length) in Refs.\ \onlinecite{Millis00b,Bartosch99d,Bartosch00d,Monien01}.

\begin{figure}[b]
\begin{center}
\epsfig{file=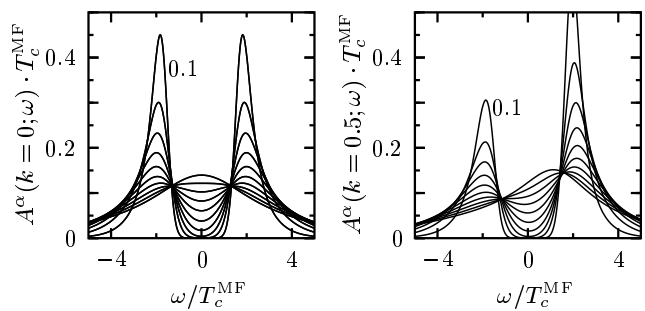} 
\end{center}
\vspace{-1mm}
\caption{Spectral function $A^{\alpha}(k;\omega)$ for $k=0$ and $k=0.5\,T_c^{\textrm{MF}}/v_F$. The temperature varies form $T/T_c^{\text{MF}}=0.1$ to $1.0$ in steps of $0.1$.}
\label{fig:SF}
\end{figure}

\begin{figure}[tb]
\begin{center}
\epsfig{file=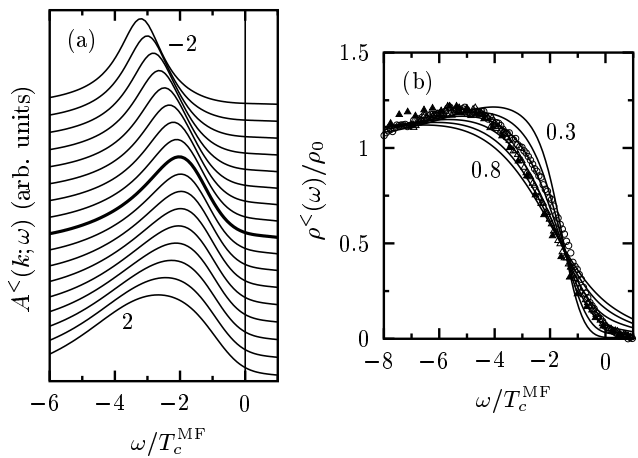} 
\end{center}
\vspace{-1mm}
\caption{(a) $A^{<}(k;\omega)$ for $T=0.5\, T_c^{\text{MF}}$ and $v_F k/T_c^{\textrm{MF}} =-2$ to $2$ in steps of $0.25$. Thick line: $k=0$, i.e.\ $p=k_F$. The spectra are normalized to the same maximum hight. (b) Solid lines: $\rho^{<}(\omega) = \int \frac{dk}{2\pi} A^{<}(k;\omega)$ for $T/T_c^{\textrm{MF}} = 0.3$ to $0.8$ in steps of $0.1$. Comparison with experimental data: Empty circles (Ref.\ \onlinecite{Gweon01}): $T=250$\,K. Empty triangles (Ref.\ \onlinecite{Gweon96}): $T=300$\,K. Filled triangles (Ref.\ \onlinecite{Dardel91}): $T=313$\,K. As a fit parameter we have used $T_c^{\textrm{MF}} = 60$\,meV$=696$\,K.}
\label{fig:ARPES}
\end{figure}

As can be seen in Fig.\ \ref{fig:SF}, at low temperatures $A^{\alpha}(k;\omega)$ exhibits two peaks instead of a quasi particle peak.
Although the qualitative features of the spectral function agree with perturbation theory or the Sadovskii approximation \cite{McKenzie96b}, on a quantitative level there are large deviations, being due to the non-Gaussian nature of the order parameter fluctuations.
To compare our results with experiments, we note that for large photon energies the conventional interpretation of the ARPES line shape predicts that it is essentially proportional to \cite{Gweon01,Joynt99}
\begin{equation}
  \label{eq:ARPESintensity}
  I(k;\omega) = \sum_{k^{\prime}\approx k} \sum_{\omega^{\prime}\approx \omega} A^{<}(k^{\prime};\omega^{\prime})\;,
\end{equation}
where $A^{<}(k;\omega) \equiv A^{\alpha}(k;\omega)/(e^{\beta \omega} +1)$ is the spectral function multiplied with the Fermi function. 
In Fig.\ \ref{fig:ARPES}(a) we show a plot of $A^{<}(k;\omega)$ for $T=0.5\, T_c^{\text{MF}}$ and different values of $k$. $A^{<}(k;\omega)$ agrees reasonably well with the ARPES spectra presented in Fig.\ 1(a) of Ref.\ \onlinecite{Perfetti02} which were taken at $T=210\,$K.
In Fig.\ \ref{fig:ARPES}(b) we compare angle integrated photoemission data with our theory. The only fit parameter used is $T_c^{\textrm{MF}} = 60$\,meV$=696$\,K. This value is less than a factor of two larger than $T_c^{\textrm{MF}} \approx 400$\,K, as obtained from optical conductivity measurements \cite{Gruener94}. The deviation might be due to a renormalization of $\lambda$ by electron-electron interactions and the simplicity of the model. 
Nonetheless, $\rho^{<}(\omega) = \int \frac{dk}{2\pi} A^{<}(k;\omega)$ does give a good fit to experimental data. We conclude that order parameter fluctuations seem to dominate the pseudogap phase.

In summary, we have calculated the spectral function of a $1d$ electronic system coupled to order parameter fluctuations determined by a generalized Ginzburg-Landau theory. Our results show that the essential physics of the pseudogap phase of quasi $1d$ charge density wave systems can be captured within a strictly $1d$ model based on order parameter fluctuations, treating their non-Gaussian statistics exactly. These order parameter fluctuations lead to non-Fermi liquid behavior and the experimentally observed absence of a Fermi edge.
\begin{acknowledgments}
The author would like to thank P.\ Kopietz, H.\ Monien, and R.\ H.\ McKenzie for discussions.
This work was partially supported by the DFG via Forschergruppe FOR 412, Project No. KO 1442/5-1.

\end{acknowledgments}



\end{document}